\documentclass[9pt,twocolumn,twoside]{opticajnl}
\journal{Optica} % use for journal or Optica Open submissions

\setboolean{shortarticle}{false}
\usepackage{lineno}
\usepackage{verbatim}
\usepackage{siunitx}
\DeclareSIUnit\dBm{dBm}
\DeclareSIUnit\dBc{dBc}

\title{A low repetition rate optical frequency comb}

\author[1,2,3,*]{Francesco Canella}
\author[2,4]{Johannes Weitenberg}
\author[2,5]{Muhammad Thariq}
\author[2,6]{Fabian Schmid}
\author[2,7]{Paras Dwivedi}
\author[3]{Gianluca Galzerano}
\author[2,7]{Theodor W. H\"ansch}
\author[2,7]{Thomas Udem}
\author[2,**]{Akira Ozawa}

\affil[1]{Dipartimento di Fisica, Politecnico di Milano, piazza Leonardo da Vinci 32, 20133 Milan, Italy}
\affil[2]{Max-Planck-Institut für Quantenoptik, Hans-Kopfermann-Straße 1, 85748 Garching, Germany}
\affil[3]{Istituto di Fotonica e Nanotecnologie - Consiglio Nazionale delle Ricerche, piazza Leonardo da Vinci 32, 20133 Milan, Italy}
\affil[4]{Fraunhofer-Institut für Lasertechnik ILT, Steinbachstraße 15, 52074 Aachen, Germany}
\affil[5]{Karlsruhe School of Optics and Photonics, Karlsruher Institut für Technologie, Schlossplatz 19, 76131 Karlsruhe, Germany}
\affil[6]{Institute for Quantum Electronics, ETH Zürich, Otto-Stern-Weg 1, 8093 Zurich, Switzerland}
\affil[7]{Fakultät für Physik, Ludwig-Maximilians-Universität München, Schellingstraße 4, 80799 Munich, Germany}
\affil[*]{francesco.canella@polimi.it}
\affil[**]{akira.ozawa@mpq.mpg.de}

\begin{abstract}
Reducing the pulse repetition rate of an optical frequency comb increases the pulse energy for a given average power. 
This enhances the efficiency of nonlinear frequency conversion and it facilitates extending the accessible wavelength range, for example into the extreme ultraviolet (XUV).
The resulting spectrally dense frequency comb can still be used for precision spectroscopy of narrow atomic or molecular transitions. %In this article, we address the question of how low the repetition rate can be before the intrinsic line broadening washes out the comb structure.
In this article, we demonstrate a low-noise infrared frequency comb with a repetition rate as low as \SI{40}{\kilo\Hz} using a Yb:KYW mode-locked laser, pulse picking, and subsequent amplification. The frequency comb structure is confirmed by generating a beat note with a continuous wave reference laser. A comb mode is actively stabilized to the reference laser, and the integrated rms phase noise from \SI{20}{\Hz} to \SI{20}{\kilo\Hz} is measured to be \SI{195}{\milli\radian}.
\end{abstract}

\setboolean{displaycopyright}{false}
\begin{document}
\maketitle

\section{Introduction}
\label{introduction}
Optical frequency combs have revolutionized the field of optical frequency metrology \cite{Udem2002, Jones2000, Diddams2020} and are indispensable tools for high-precision laser spectroscopy to study fundamental physics \cite{Beyer2017, Grinin2020, Patra2020} and for optical frequency standards \cite{Diddams2001, Bloom2014, Ushijima2015, Schioppo2017, Brewer2019}.
The first application of frequency combs was to measure the frequency of a continuous wave laser that is then used for precision spectroscopy \cite{Udem1999, Holzwarth2001, Diddams2020}. It is also possible to use the frequency comb itself to excite target transitions and perform spectroscopy with ultra-wide spectral coverage, fast detection times, and high sensitivities \cite{Grinin2020, Thorpe2006, Diddams2007, Gohle2007, Coddington2008, Picque2019}.

An optical frequency comb consists of many spectral modes that are equally spaced by the repetition rate of the generating mode-locked pulse train which is typically in the \SI{}{\MHz} to the \SI{}{\GHz} range for conventional solid-state and fiber laser-based oscillator designs.
While combs with mode spacings of several tens of GHz have found applications for calibration of astronomical spectrographs \cite{Steinmetz2008, Li2008}, %Even higher repetition rates exceeding one hundred GHz have been demonstrated with ultrahigh $Q$ whispering gallery mode microresonators \cite{Marin-Palomo2017, Fujii2020}.
a pulse train with a relatively low repetition rate below \SI{10}{\MHz} can have correspondingly higher pulse energy and is therefore advantageous for several applications, including for example driving efficient nonlinear processes \cite{Masters2004, Hadrich2015, Kottig2020}. % and material processing \cite{Eaton2005, Gattass2008}.
Nonlinear frequency conversion of optical frequency combs can enable precision spectroscopy in wavelength ranges where continuous-wave lasers are not available, such as the extreme ultraviolet (XUV) \cite{Cingoz2012, Schliesser2012, Ozawa2013}. For example, our planned experiment of precision spectroscopy of the 1S-2S transition in He$^{+}$ ions requires an optical frequency comb at 60.8 nm \cite{Herrmann2009, Moreno2023}. The Ramsey-type frequency comb, which consists of pairs of intense pulses, can be an alternative method to address transitions at XUV wavelengths \cite{Dreissen2019}.
Precision spectroscopy of the nuclear transition of $^{229}$Th at around 149 nm may find application as a nuclear optical clock \cite{VonderWense2020, Peik2021, Zhang2022, Kraemer2022}.

Intra-cavity high-order harmonic generation allows the generation of high-power XUV frequency combs suitable for direct frequency comb spectroscopy  \cite{Gohle2005, Jones2005, Ozawa2008, Cingoz2012, Pupeza2021, Seres2019}. In this scheme, the high-order harmonic generation process is performed inside an enhancement cavity using a gaseous or solid medium placed at the focus of the cavity. Special care must be taken to avoid detrimental effects of the high average power and intensity, such as thermal lensing \cite{Carstens2014}, plasma phase shift \cite{Holzberger2015, Carlson2011, Allison2011}, misalignment due to elevated temperature \cite{Carstens2014}, and damage to the optics \cite{Wang2020}. In addition, enhancement cavities for ultrashort pulses have to be carefully designed to tailor the intra-cavity dispersion \cite{Schliesser2006, Lilienfein2017}.
%and to minimize the frequency-to-intensity noise conversion \cite{Morville2002, Canella2022}.

Alternatively, by operating a comb at a lower repetition rate, a similarly high pulse energy could be achieved with modest average power even without an enhancement cavity.
Lowering the repetition rate results in a frequency comb with a smaller mode spacing. For precision spectroscopy, the mode spacing should be at least several times larger than the linewidth of the transition under investigation in order to obtain a comb-mode resolved spectroscopy signal. 
For instance, a signal linewidth of about \SI{1}{\kHz} is expected for the 1S-2S transition in He$^{+}$ at 60.8~nm \cite{Herrmann2009, Moreno2023}, while the nuclear transition of $^{229}$Th at about 149 nm has a natural linewidth of about \SI{20}{\micro\Hz} \cite{VonderWense2020, Peik2021}. 
In principle, a comb with a few kHz mode spacing would be sufficient for these applications.

High pulse energy ultrafast lasers at such low repetition rates are conventionally generated using master oscillator power amplifiers (MOPAs), where the chirped pulse amplification (CPA) scheme is employed and have found various applications including attosecond physics \cite{Krausz2009, Sansone2006}, laser particle acceleration \cite{Lundh2011, Gonsalves2019}, and ultrafast transient spectroscopy \cite{Polli2010}. % and investigation of ultrafast electron emission process from solids \cite{Kruger2011}. 
By actively controlling the round-trip phase-shift of the oscillator, each pulse can have an identical carrier-envelope phase (CEP), which is particularly important for the study of field-sensitive phenomena driven by few-cycle pulses \cite{Baltuska2003}.
Using the balanced optical cross-correlator method \cite{Schibli2003}, the pulse-to-pulse timing jitter of the CPA laser system can be controlled precisely. For example, in Ref. \cite{Huang2011} a sub-100~fs timing jitter is demonstrated at \SI{1}{\kHz} repetition rate. A pulse timing jitter of 100~fs still introduces a relative frequency uncertainty of 10$^{-10}$ at a repetition rate of \SI{1}{\kHz}. This corresponds to a broadening of the optical comb modes that is larger than the mode spacing and could wash out the comb structure in the frequency domain.  Although such CEP-stabilized low-repetition-rate laser systems have been shown to be suitable for studying ultrafast phenomena, they do not guarantee a low-noise frequency comb structure.

In this work, we report a low-noise optical frequency comb that operates at tunable repetition rates from 40 kHz to 40 MHz using a Yb:KYW mode-locked oscillator. Mode-locked lasers oscillating directly at sub-MHz repetition rates would require very long cavities which may not be practical. Instead, conventional mode-locked lasers and pulse pickers are used to generate optical frequency combs at low repetition rates \cite{Khwaja2020, Saule2017}.
The associated power loss is compensated by re-amplifying the pulse train such that higher pulse energy is achieved. A comb mode is actively stabilized to an ultra-stable continuous wave (cw) reference laser. The phase noise of the stabilized mode is characterized with respect to the reference laser and is shown to result in a narrow linewidth suitable for exciting narrow transitions.

%%%%%%%%%%%%%%

\section{Modelling of the pulse-picking process}
\label{theory}
In this section, we first model the pulse picking process by treating the pulse picker as an ideal amplitude modulator.
%We then introduce the effect of the frequency shift when an acousto-optic modulator-based pulse picker is used.

For mathematical convenience, we model the output of the mode-locked laser as a train of Gaussian-shaped pulses with a repetition rate $f_{\rm rep}$. The temporal pulse spacing is $T = f_{\rm rep}^{-1}$. The electric field $E(t)$ of the laser pulse train can be described as
\begin{equation}
E(t) = A \sum_{k=-\infty}^{\infty}  \exp{\left[ - \frac{\left( t - k\, T\right)^2}{\Delta t^2}  \right]} e^{i \left(\omega_0 t+\varphi\left( t \right)\right)},
\label{electric_field_1}
\end{equation}
%%
%%%%%%%%%
\begin{figure*}[h!]
\centering\includegraphics[width=\textwidth]{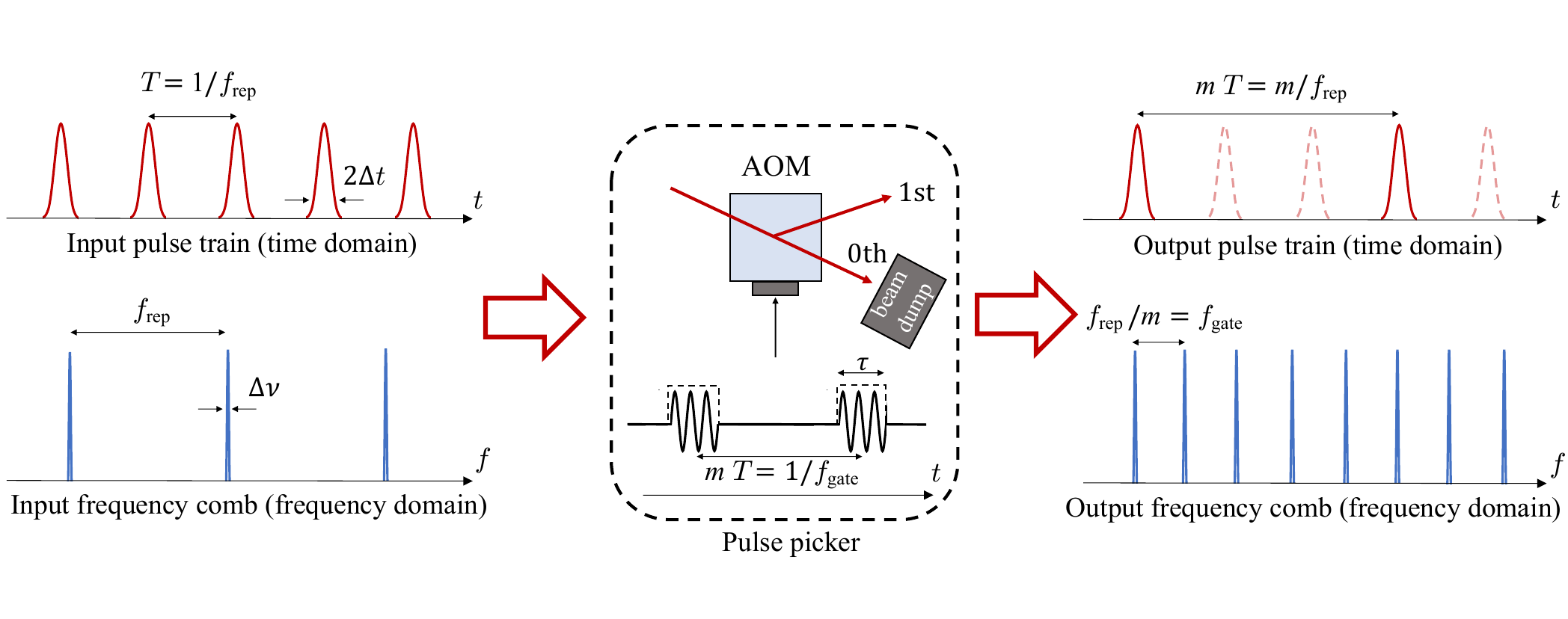}
\caption{Low-repetition rate frequency comb generation using an AOM-based pulse picker. 
Every $m$-th pulse is diffracted by the AOM to the 1st diffraction order, while the others remain in the 0th order and are dumped. After the pulse picking, the pulse-to-pulse time interval increases by the factor $m$, while the mode spacing in the frequency domain becomes $m$ times smaller. The AOM is driven by an RF carrier which is amplitude modulated with a rectangular-shaped gate-pulse train.
}
\label{pulsepickingscheme}
\end{figure*}
%%%%%%%%
where $A$ is the field envelope peak amplitude, $\omega_0$ is the angular frequency of the carrier, and $\Delta t$ is the pulse duration defined by the $1/e$ half-width of the field amplitude. The residual phase fluctuation after comb stabilization is represented by $\varphi( t )$.
For simplicity, the CEO frequency of the comb is assumed to be zero in eq.(\ref{pulsepickingscheme}) by assuming that the carrier frequency $\omega_0$ is an integer multiple of the pulse repetition rate $2\pi/T$.  Including a finite CEO frequency in the calculation is straightforward, and does not change the results.
We introduce pulse picking by assuming that the laser pulses pass through an ideal amplitude modulator with rectangular-shaped gating which selects every $m$-th pulse. We call $m$ the pulse picking factor.
The pulse picker reduces the average power and the repetition rate by a factor $m$ and hence the power in each of the modes by a factor $m^2$.
Under the approximation that the electric field amplitude at the rising and falling edge of the rectangular-shaped gating is negligible, the pulse-picked field $E_{\rm p}(t)$ can be written as 
\begin{equation}\begin{split}
E_{\rm p}(t) &= A \sum_{k=-\infty}^{\infty}  \exp{\left[ - \frac{\left( t - m \, k\, T\right)^2}{\Delta t^2}  \right]} e^{i \left(\omega_0 t+\varphi\left( t \right)\right)} \\
& \equiv E_{\rm p, 0}(t) e^{i \varphi\left( t \right)},
\label{electric_field_2}
\end{split}
\end{equation}
%%
%where we assumed that $N$ is an integer multiple of $m$ and defined $E_{\rm p, 0}(t)$ as the noiseless component of the pulse-picked field. The total number of pulses, and therefore the average power of the pulse train, is reduced by a factor of $m$.
where  $E_{\rm p, 0}(t)$ is defined as the noiseless component of the pulse-picked field.
The spectrum of the pulse-picked field $E_{\rm p}(t)$ is given by
\begin{equation}
\tilde{E}_{\rm p}(\omega) = \int_{-\infty}^{\infty} E_{\rm p}(t) e^{-i\omega t} {\rm d}t \approx \tilde{E}_{\rm p, 0}(\omega) + \frac{i}{2 \pi}\, \left(\tilde{E}_{\mathrm{p},0} \ast \tilde{\varphi} \right)(\omega),%\equiv E_{\rm p, 0}(\omega) + E_n(\omega)
\label{electric_field_3}
\end{equation}
where the approximation is valid for small rms phase noise with a vanishing mean, i.e. $e^{i\varphi\left(t\right)} \approx 1+ i\, \varphi\left(t\right)$.
In Eq. (\ref{electric_field_3}), the convolution is defined as \mbox{$\left(\tilde{E}_{\mathrm{p},0} \ast \tilde{\varphi} \right)(\omega) = \int_{-\infty}^{\infty} \tilde{E}_{\rm p, 0}(\omega^\prime)\tilde{\varphi}(\omega-\omega^\prime) {\rm d}\omega^\prime$}, with $\tilde{E}_{\rm p, 0}(\omega)$ and $\tilde{\varphi}(\omega)$ are Fourier transforms of $E_{\rm p, 0}(t)$ and $\varphi(t)$, respectively. The main Fourier components of $\tilde{E}_{\rm p,0}(\omega)$ and $\tilde{\varphi}(\omega)$ are in the optical and the radio frequency domain respectively. The spectrum of the noiseless component is given by:
\begin{align}
\tilde{E}_{\rm p, 0}(\omega) = &  \,A\,\!\!\int_{-\infty}^{\infty}\!\sum_{k=-\infty}^{\infty}\!\!\exp{\left[ - \frac{\left( t - m \, k\, T\right)^2}{\Delta t^2}  \right]} e^{i (\omega_0 -\omega)t} {\rm d}t \nonumber \\ \nonumber
= & \frac{2\pi ^{3/2}\Delta t}{mT} A  \exp{\left[ -\frac{1}{4}\Delta t^2 \left( \omega - \omega _0 \right)^2 \right]} \\ 
& \times \sum_{n = -\infty}^{\infty} \delta \left( \omega - \frac{2\pi n}{mT} \right) \label{electric_field_4}.
\end{align}

As expected, the spectrum after pulse picking consists of comb modes with a mode spacing of $1/mT= f_{\rm rep}/m$, as expressed by the sum over the delta functions $\delta(\omega-2\pi n/mT)$. The newly created comb modes can be considered as the sidebands at subharmonics of the original repetition rate introduced by the amplitude modulation that selects every $m$-th pulse (see Supplement for a detailed derivation). These additional sidebands fill in the gaps between the modes of the original comb. The Gaussian spectral envelope is maintained (in the limit of $\Delta t \ll \tau$), also for the newly added modes, since the picked pulses are still Gaussian in the time domain.  

The spectral intensity can be calculated as $\tilde{I}_{\rm p, 0}(\omega) \propto |\tilde{E}_{\rm p, 0}(\omega)|^2$ demonstrating that the comb-mode power scales with $m^{-2}$ as explained above. Note that $|\tilde{E}_{\rm p, 0}(\omega)|^2$ includes the squares of the delta functions, i.e. infinite power density at the frequencies of the modes compatible with an infinite number of pulses. 
%This is consistent with the assumption of an infinite number of pulses ($N \to \infty$), which leads to an infinite power density at the frequencies of the modes.
Using Eq. (\ref{electric_field_4}) in the last term of Eq. (\ref{electric_field_3}), we find that the phase noise is convolved into all comb modes, including the new modes created by pulse picking. 
The phase noise spectrum $\tilde{\varphi}(\omega)$ for radio frequencies $|\omega|>2\pi /mT$ is folded into $|\omega|<2\pi /mT$. In the time domain, the effect can be considered as an aliasing where the phase noise at $|\omega|>2\pi /mT$ is undersampled by the pulse train.
If the phase noise spectrum $\tilde{\varphi}(\omega)$ is flat and extends up to the original repetition frequency, reducing the repetition rate by a factor of $m$ will result in a $m$-fold increase in the power spectral density (PSD) of the phase noise in the frequency range of $|\omega|<2\pi /mT$.
On the other hand, the integrated phase noise does not change after pulse picking because the $m$-fold increase in the PSD is canceled by an inverse reduction of the maximum frequency, i.e. the Nyquist frequency reduces by a factor $m$. This is consistent with Eqs. (\ref{electric_field_1}) and (\ref{electric_field_2}) which contain identical noise terms $e^{i \varphi\left( t \right)}$, and the rms phase noise is expected to be identical before and after pulse picking.

The model described here shows that the phase noise already present in the original pulse train affects the frequency comb structure equally before and after pulse picking, regardless of the repetition rate. A trivial requirement is that the linewidth of the comb modes prior to pulse picking should be narrower than the mode spacing after pulse-picking to avoid washing out the comb structure. 
In addition, the actual implementation of the pulse picker and the subsequent amplification should be low-noise to preserve the comb structure at lower repetition rates. Our work described in sections~\ref{experiment} and \ref{results} aims to experimentally confirm that this is possible.

Pulse picking can be implemented using an acousto-optical modulator (AOM) or a combination of an electro-optic modulator (EOM) and a polarizer. 
In Fig. \ref{pulsepickingscheme}, we show a conceptual scheme of an AOM-based pulse picking. The AOM is driven at a carrier frequency $f_{\rm AOM}$ (typically tens or hundreds of \SI{}{\MHz}) which is amplitude modulated with gating pulses.
Here we assume a rectangular-shaped gate function with repetition rate $f_{\rm gate}$ (time spacing $T_{\rm gate}=f_{\rm gate}^{-1}$) and a width of $\tau$.\footnote{In the Supplement, we discuss the effect of gating timing jitter.}
The RF drive signal for the AOM can be described as 
\begin{equation}
V_{\rm RF}(t) = \sum_{n=-\infty}^{\infty} A_{\rm RF}\left( t - n T_{\rm gate}\right) e^{i\omega_{\rm AOM} t},
\label{AOM_field}
\end{equation}
where $A_{\rm RF}$ is the rectangular-shaped gate function, and $\omega_{\rm AOM} = 2\pi f_{\rm AOM}$.\\
The modulator diffracts every $m$-th pulse to the 1st diffraction order, and the rest is sent to a beam dump. 
The Fourier transform of Eq. (\ref{AOM_field}) shows that the spectrum of the RF signal consists of narrow lines spaced by $f_{\rm gate}$, similar to an optical frequency comb.
The frequency of the RF modes is given by
\begin{equation}
f_{{\rm RF}, n'} =  n^\prime \,f_{\rm gate}+f_{\rm AOM},
\label{RF_frequency}
\end{equation}
with an integer $n^\prime$.
The frequency of the optical comb modes $f_{n,n^\prime}$ after pulse picking is given by the sum of the original comb mode frequencies and the AOM RF frequencies:
\begin{align}
f_{n,n^\prime} & = n\, f_{\rm rep} +f_{\rm CEO}+ f_{{\rm RF}, n'}
\\ \nonumber
& = n\, f_{\rm rep} + n^\prime \,f_{\rm gate}+f_{\rm CEO}+f_{\rm AOM},
\label{comb_pulse_picking}
\end{align}
where $f_{\rm CEO}$ is the CEO frequency of the original comb.
To obtain an equidistant comb structure after pulse-picking, the ratio $f_{\rm rep}/f_{\rm gate}$ must be an integer. In the time domain, this condition translates to the requirement that $T_{\rm gate} = m \, T$, with $m$ introduced in Eq.~(\ref{electric_field_2}).
The CEO of the pulse-picked comb remains unchanged after pulse picking if $f_{\rm AOM}/f_{\rm gate} = q$ is an integer.
With an integer $q$, we can define a new mode-index $\tilde{n}\equiv n^\prime +q+n\, m$ to find a compact expression for the mode frequencies of the pulse-picked comb
\begin{equation}
f_{\tilde{n}}= \tilde{n} \, \frac{f_{\rm rep}}{m} + f_{\rm CEO}.
\label{low_rep_rate_modes}
\end{equation}
In the special case of zero CEO frequency ($f_{\rm CEO}=0$), a pulse train with a constant CEP can be obtained which finds interesting applications in attosecond physics~\cite{Ishii2014, DeVries2015, Calegari2016}. 

%%%%%%%%%%%%%%

\section{Experiment}
\label{experiment}
%%%%%%%%
\begin{figure*}[h]
\centering\includegraphics[width=\textwidth]{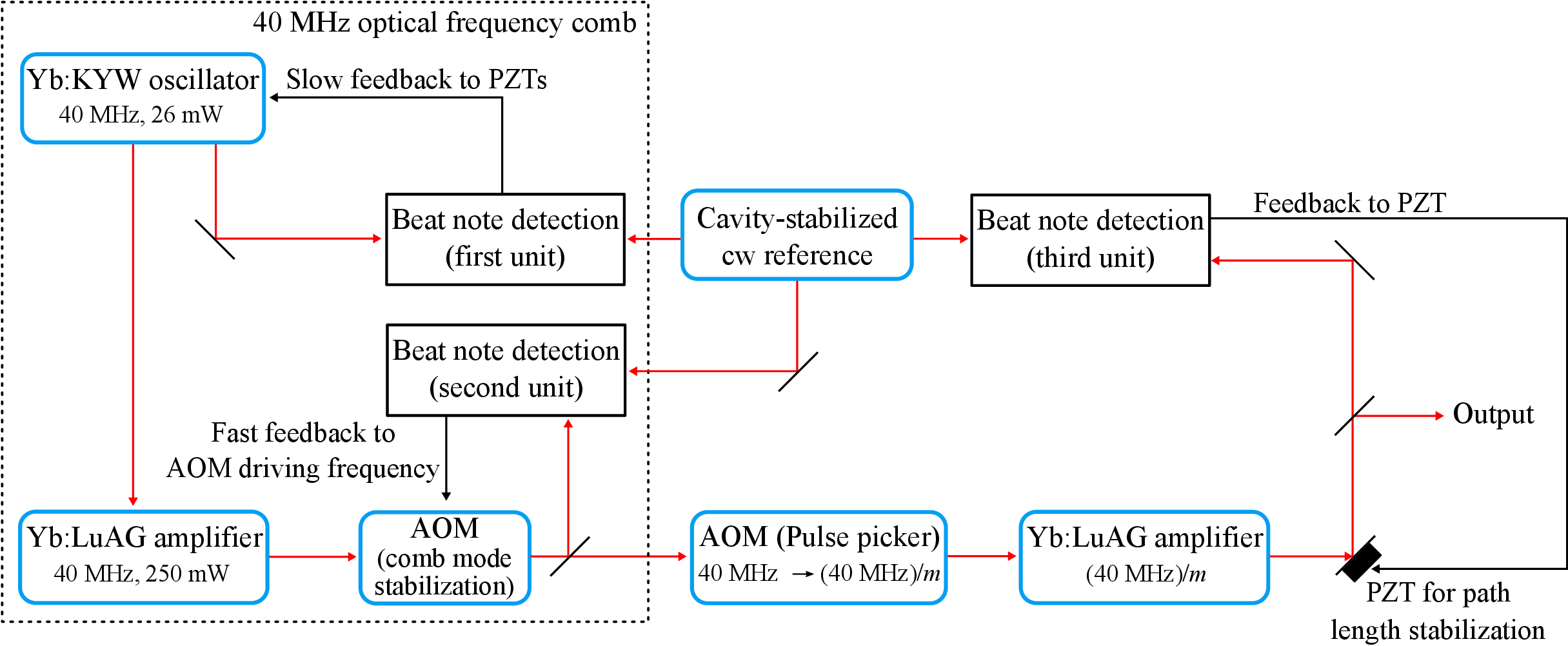}
\caption{Schematic of the experimental setup.  The 40 MHz optical frequency comb (dotted box) consists of a mode-locked Yb:KYW oscillator, a solid-state amplifier using Yb:LuAG as gain medium, and an AOM frequency shifter for fast control of the stabilized comb mode frequency. Slower control acts on PZT-actuated mirrors of the laser cavity (PZTs). The output of the Yb:KYW oscillator is centered at \SI{1030}{\nm} with a bandwidth of \SI{14}{\nm}. The error signal for phase stabilization of one of the comb modes is obtained from beat notes between that mode and an ultra-stable cw reference laser emitting at 1033~nm. The cw reference laser is amplified by a semiconductor optical amplifier (BOA1050P Thorlabs, not shown) before sending it to the second and third beat note detection units. An AOM-based pulse picker reduces the comb's repetition rate to $(\SI{40}{\MHz})/m$, where $m$ is the pulse picking factor. After the pulse picker, the pulses are re-amplified by a second Yb:LuAG amplifier. The third feedback loop controls a PZT-actuated mirror in the beamline and reduces the phase noise due to fluctuations in the beam path length.}
\label{experimental_setup}
\end{figure*}
%%%%%%%%
Figure \ref{experimental_setup} shows our setup for generating and testing a low repetition rate optical frequency comb.
A home-built Yb:KYW oscillator is mode-locked by soft-aperture Kerr-lensing and generates a \SI{40}{\MHz} pulse train. The output spectrum is centered at \SI{1030}{\nm} and has a FWHM bandwidth of \SI{14}{\nm}. The average output power is \SI{26}{\milli\watt}. The FWHM temporal pulse duration is measured to be \SI{89}{\femto\s} using the intensity autocorrelation method (autocorrelation length \SI{137}{\femto\s}), assuming sech$^2$ pulse shape.

Two PZT-actuated mirrors are installed in the laser cavity to control the cavity length and are used to stabilize the frequency comb (PZT stands for lead zirconate titanate). One has a bandwidth of about \SI{10}{\kHz}, while a second is used to compensate for slow drifts. The laser cavity is placed on a vibration-isolated, temperature-stabilized aluminum baseplate and installed inside an air-tight aluminum housing \cite{Schmid2023}.

At the laser output, the repetition rate $f_{\rm rep}$ is detected with a fast photodiode (Thorlabs DET01CFC, not shown in Fig.~\ref{experimental_setup}).
The laser has an auxiliary output that is taken from the reflection of an intracavity optic element with a compromised spectral phase compared with the main output. This second output has about \SI{60}{\milli\watt} of power and is sent into a heterodyne beat detection setup with a continuous wave (cw) reference laser. The cw reference laser operates at \SI{1033}{\nano\m} and is stabilized to an ultra-stable reference cavity. The rms phase noise of the reference laser was measured to be 10.2 mrad integrated from \SI{10}{\kHz} to \SI{10}{\MHz} with respect to the reference cavity \cite{Schmid2019}.

In the heterodyne beat detection setup, about 100 comb modes are filtered out around the frequency of the cw reference laser using an interference filter (Alluxa A4017) and an etalon (LightMachinery OP-6204-M, \SI{7.3}{\giga\hertz} FWHM bandwidth).
The beat signal between the frequency comb and the reference laser is detected using balanced photodetectors (Koheron PD100B) which suppresses the contribution of classical amplitude noise \cite{Carleton1968}. The RF beat signal is filtered to isolate the beat note between the closest comb modes and the reference laser. Then the signal is phase-compared to a \SI{10}{\MHz} signal generated by a signal generator (Marconi Instruments 2022C). The signal controls the PZT actuators of the Yb:KYW laser via a home-built loop filter. This way, one of the comb modes is phase-stabilized to the cw reference laser.

%The laser output passed through a Faraday isolator and mode-matching optics before it was amplified by a solid-state double-pass Yb:LuAG amplifier pumped by a multimode diode laser operated at \SI{932}{\nano\m}.
The frequency comb is amplified by a solid-state double-pass Yb:LuAG amplifier pumped by a multimode diode laser operating at around \SI{935}{\nano\m}.
With an input seed power of \SI{26}{\milli\watt}, we obtain \SI{250}{\milli\watt} at the output when the pump power is set to \SI{7.2}{\watt}. Since the gain bandwidth of Yb:LuAG is about \SI{5}{\nano\m} \cite{Siebold2012}, a significant gain narrowing effect reduces the bandwidth of the amplifier output to \SI{2.7}{\nano\m}. The peak gain at \SI{1030}{\nano\m} is approximately \SI{16.6}{\decibel}.
The amplifier output is sent to an AOM (AA Opto Electronic MT110) that is used to stabilize one of the comb modes in combination with the PZTs in the laser cavity. %to control the CEO frequency.
The first-order diffraction of the AOM downshifts the entire frequency comb of the laser by about \SI{110}{\MHz}. The diffraction efficiency of the AOM is about \SI{70}{\percent}. The second beat signal between the frequency comb and the reference cw laser is obtained after the AOM. The beat signal is compared with the \SI{10}{\MHz} frequency reference and is sent to a loop filter (Vescent Photonics D2-125). The loop filter's output is sent to a voltage-controlled oscillator (VCO, Pasternack Enterprises Inc. PE1V31008) which generates the RF signal that drives the AOM.
The control bandwidth is estimated to be >\SI{100}{\kHz}, limited by the time required for the acoustic wave inside the AOM to reach the laser beam.
%%%%%%%%
\begin{figure*}[h]
\centering\includegraphics[width=0.8\textwidth]{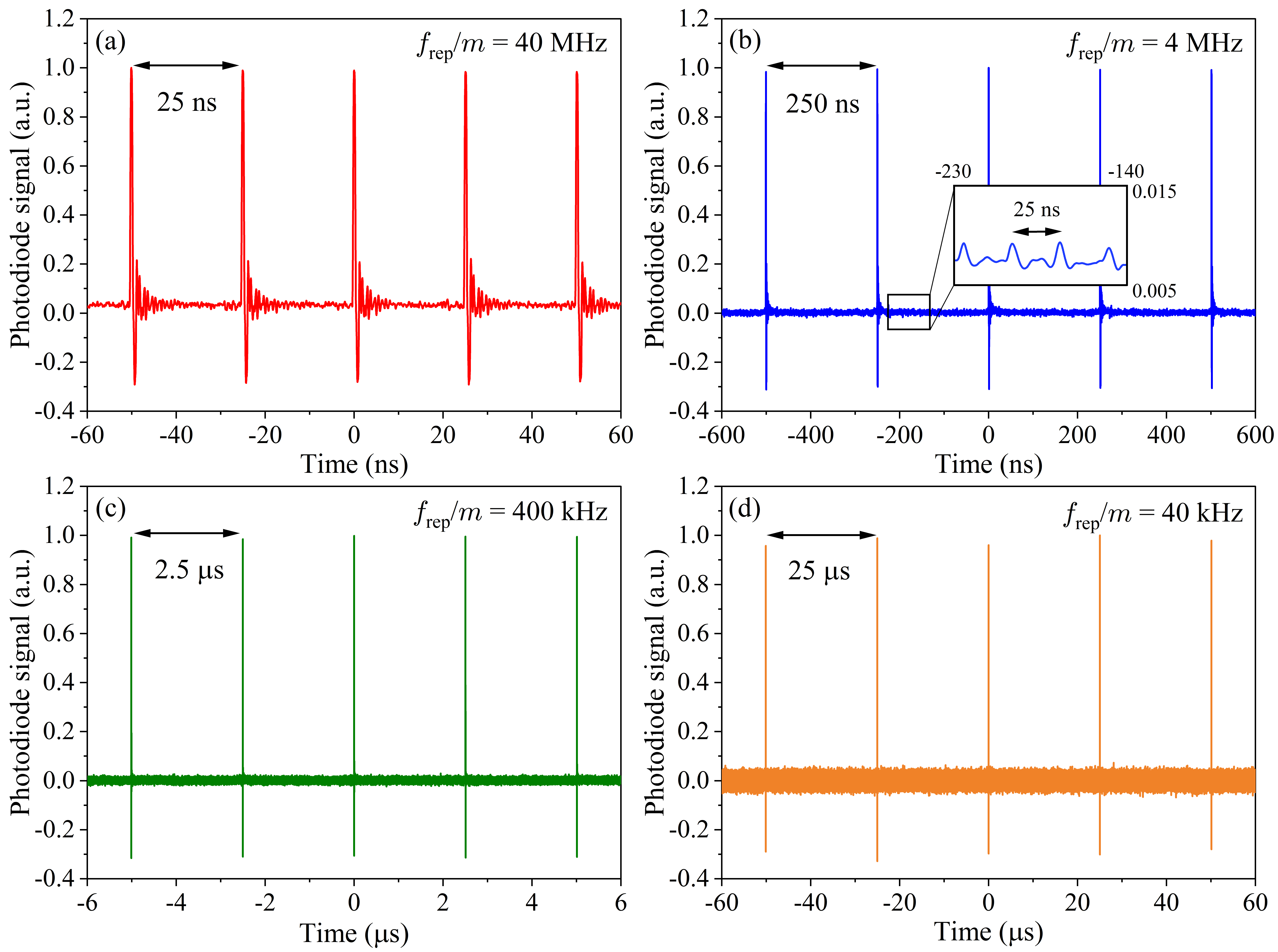}
\caption{
Time domain traces of pulses after pulse picking and the second amplifier ("Output" in Fig. \ref{experimental_setup}). The negative signal is due to ringing. a) At a repetition rate of \SI{40}{\mega\hertz} without pulse picking ($m=1$). b) At a repetition rate of \SI{4}{\mega\hertz} which corresponds to a pulse-to-pulse interval of \SI{250}{\nano\s} and a pulse picking factor of $m=10$. The magnified inset shows a trace averaged over 1000 acquisitions where residual \SI{40}{\MHz} pulses are visible with about \SI{28}{\dB} of supression. c) Pulses at a repetition rate of \SI{400}{\kilo\hertz}, corresponding to $m=10^2$ (\SI{2. 5}{\micro\s} pulse-to-pulse interval). d) Pulses at \SI{40}{\kilo\hertz} repetition rate, corresponding to a pulse-to-pulse interval of \SI{25}{\micro\s} and $m=10^3$. The oscilloscope sampling rate is 40 Gs/s for all traces.}
\label{timedomain}
\end{figure*}
%%%%%%%%

The following pulse picker AOM (AA Opto Electronic MT200-A0.4-1064) is driven at a carrier frequency of $f_{\rm AOM}=\SI{200}{\MHz}$. It selects every $m$-th pulse by amplitude modulating the carrier with a rectangular-shaped envelope.
A rectangular gate signal with a pulse width of $\tau=\SI{32}{\nano\s}$ at a frequency of $f_{\rm gate}=f_{\rm rep}/m$ is used. The gate pulse width is less than $2T$ as required to select individual pulses. An RF switch (Minicircuit ZASWA-2-50DR+) is used for the modulation. The diffraction efficiency of the AOM is measured to be $>$\SI{60}{\percent}, and the pulses remaining in the 0th order are sent to a beam dump.
The gate signal is generated by a delay generator (Alphanov Tombak) using the repetition rate signal from the Yb:KYW oscillator as the timing source. The RF carrier signal to drive the AOM is derived from the 5th harmonic of the repetition rate which is generated in the detection with a high-bandwidth photodiode.
A band-pass filter with a \SI{3}{\decibel}-bandwidth of \SI{10}{\MHz} is used to isolate the 5th harmonic. In this way, the gate signal, the repetition rate, and the AOM RF carrier are phase synchronized. The RF switch produces a modulation with 10-90\% rise/fall times of \SI{5}{\nano\s}.
Measuring the energy of the picked pulses for different gating delays with respect to the pulses reveals the time response of the AOM. The 10-\SI{90}{\percent} rise and fall time of the AOM was measured to be \SI{7.5}{\nano\s}. This is dominated by the time it takes for the acoustic wave to cross the focused laser beam at the point of interaction.
From the speed of sound within the modulator material ($\rm TeO_2$) of \SI{4200}{\meter/\s} and the laser beam diameter of $2w_0 = \SI{39}{\micro\m}$, the rise/fall time is estimated to \SI{6}{\nano\s}.

The picked pulses are sent to a second Yb:LuAG amplifier which is similar in design to the first amplifier. When the pump power is set to \SI{14}{\watt}, the amplifier output power is \SI{2. 5}{\watt}, \SI{374}{\milli\watt}, \SI{43}{\milli\watt}, and \SI{8}{\milli\watt} at a repetition rate of \SI{40}{\MHz}, \SI{4}{\MHz}, \SI{400}{\kHz}, and \SI{40}{\kHz}, respectively.% for picking factors of $m = 1, 10, 10^2, 10^3$.

The output pulse train of the second amplifier is measured by a fast InGaAs photodiode (Thorlabs DET01CFC, \SI{1.2}{\GHz} bandwidth) and a 2.5 GHz oscilloscope (LeCroy WavePro 7Zi).  The results are shown in Figure \ref{timedomain} at repetition rates of \SI{40}{\MHz}, \SI{4}{\MHz}, \SI{400}{\kHz}, and \SI{40}{\kHz}.

We find that the pulse-picked beam still contains a tiny fraction of the pulse train at the original repetition rate of \SI{40}{\MHz}.  This is not due to incomplete suppression of the RF carrier, but is caused by scattering within the AOM material.
We could suppress the optical power of this component by \SI{28}{\dB} compared to the picked pulses. This was achieved by carefully adjusting the size and position of an iris surrounding the pulse-picked laser beam.

A PZT-actuated mirror is introduced in the beamline after the second amplifier to compensate for possible low-frequency phase fluctuations due to free-space parts of the setup, e.g. mirrors or breadboard vibrations. In addition, the phase noise introduced by the pulse picker and amplifier is partially compensated.
A portion of the beam after the PZT actuated mirror is sent to the third beat detection setup. The beat signal is then phase compared to a reference at \SI{11.3}{\MHz} from an electronic synthesizer (Marconi Instruments 2022C) and is used as an error signal to drive the PZT via a home-built loop filter. The in-loop error signal shows a peak at about \SI{4}{\kHz} when the feedback gain is too high, indicating a control bandwidth of approximately \SI{4}{\kHz}. This is fast enough to significantly suppress low-frequency phase noise caused by mechanical and acoustic vibrations.
\raggedbottom
%%%%%%%%%%%%%

\section{Results and discussion}
\label{results}

%%%%%%%%
\begin{figure*}[h]
\centering\includegraphics[width=0.8\textwidth]{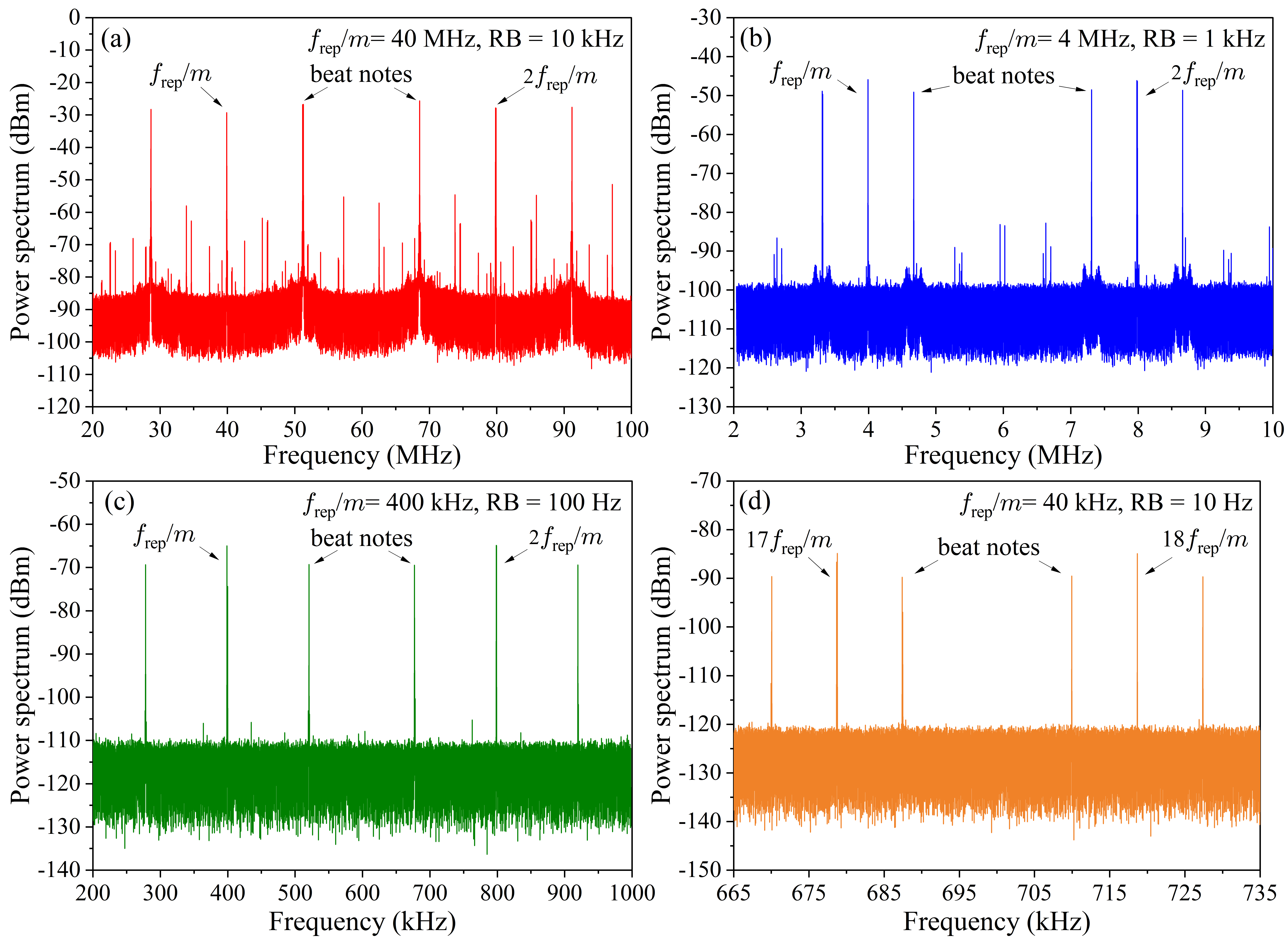}
\caption{Beat note spectra acquired from the third beat note detection unit at different repetition rates of (a) \SI{40}{\MHz}, (b) \SI{4}{\MHz}, (c) \SI{400}{\kHz}, and (d) \SI{40}{\kHz}. A resolution bandwidth (RB) of \SI{10}{\kHz}, \SI{1}{\kHz}, \SI{100}{\Hz}, and \SI{10}{\Hz} was used, respectively. 
In all plots except for panel (d), two peaks which correspond to $f_{\rm rep}/m$ and $2 f_\mathrm{rep}/m$ are visible in addition to two peaks that correspond to the beat frequencies.
The repetition rate signal is strongly suppressed by balanced photodetection.
The trace acquired for \SI{40}{\kHz} shows frequencies between the 17th and the 18th harmonics of $f_{\rm rep}$ to avoid the elevated noise floor of the measurement setup at low frequencies.
The RF spectrum at \SI{40}{\MHz} repetition rate without pulse picking shown in (a) contains several peaks between the repetition rate peaks and the beat notes. Most of these are due to the mixing of the strong repetition rate and beat note signals on the photodiode.
} 
\label{largespanbeat}
\end{figure*}
%%%%%%%%
The evaluated beat note spectra are obtained from the third beat note detection unit (Fig. \ref{experimental_setup}) and shown in Fig. \ref{largespanbeat}. 
The acquisitions were made at repetition rates of \SI{40}{\MHz} (no pulse picking), \SI{4}{\MHz}, \SI{400}{\kHz}, and \SI{40}{\kHz} with resolution bandwidths of \SI{10}{\kHz}, \SI{1}{\kHz}, \SI{100}{\Hz}, and \SI{10}{\Hz}, respectively. 
A clear comb structure is maintained even at the reduced repetition rate of \SI{40}{\kHz}, where a narrow beat note peak is still visible with a signal-to-noise ratio above \SI{30}{\dB}.
For all repetition rates, the linewidth of the beat signal is limited by the resolution bandwidth of the spectrum analyzer (Agilent E4445A).

%%%%%%%%
\begin{figure}[h]
\centering\includegraphics[width=\columnwidth]{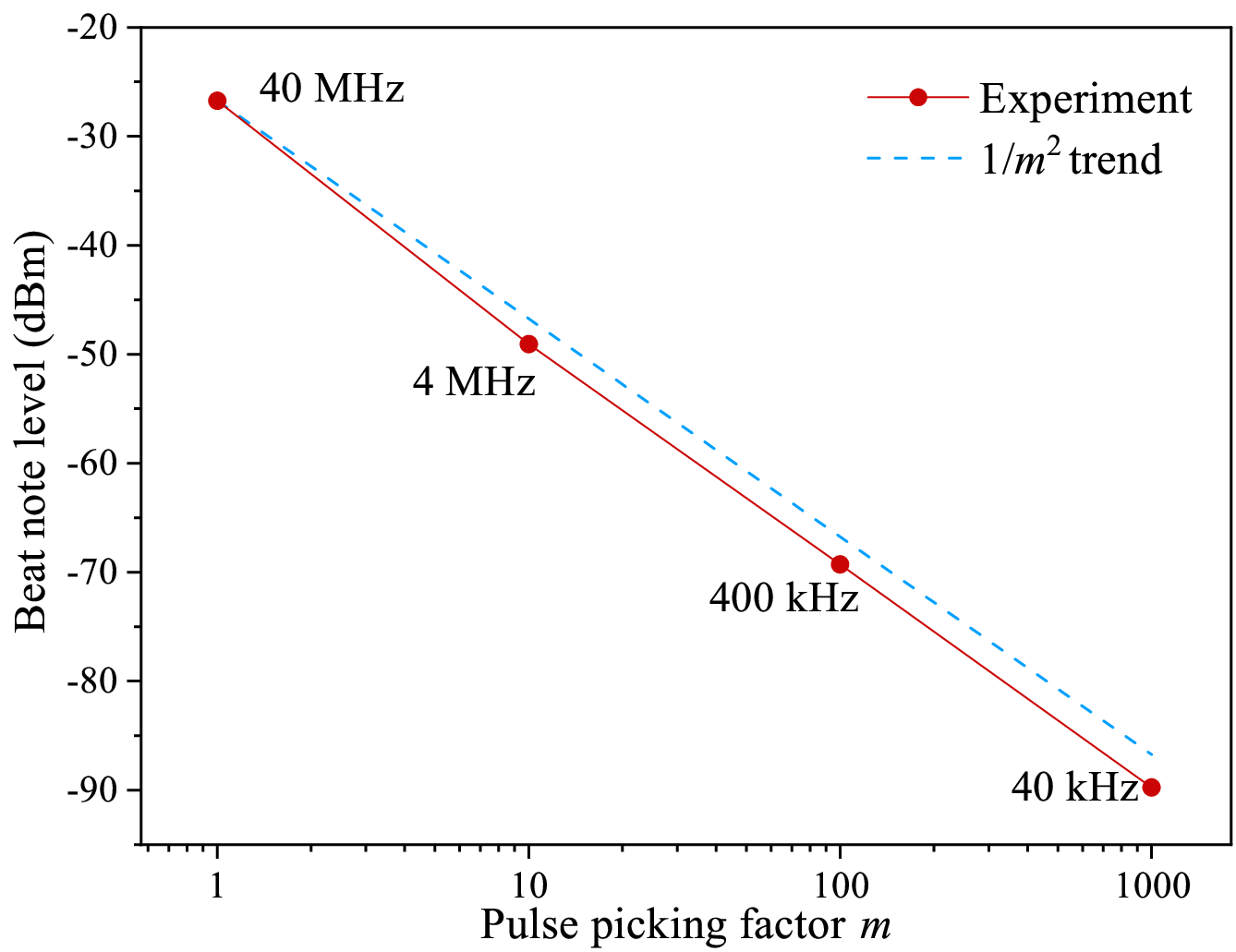}
\caption{Beat note power level as a function of the pulse picking factor $m$.
The dashed line shows a trend proportional to $m^{-2}$, as expected from the pulse picking model. The trend line is drawn assuming the \SI{40}{\MHz} point as reference.} 
\label{beat_note_power}
\end{figure}
%%%%%%%%
Fig. \ref{beat_note_power} shows the RF power of the beat notes for different pulse picking factors $m$. 
The electrical beat note power is proportional to the optical power contained in a single comb mode and scales with $1/m^2$, as expected from the spectral intensity $I_{\rm p, 0}(\omega)$ of our model in Eq. (\ref{electric_field_4}).

%%%%%%%%%%%
\begin{figure}[h]
\centering
\includegraphics[width=\columnwidth]{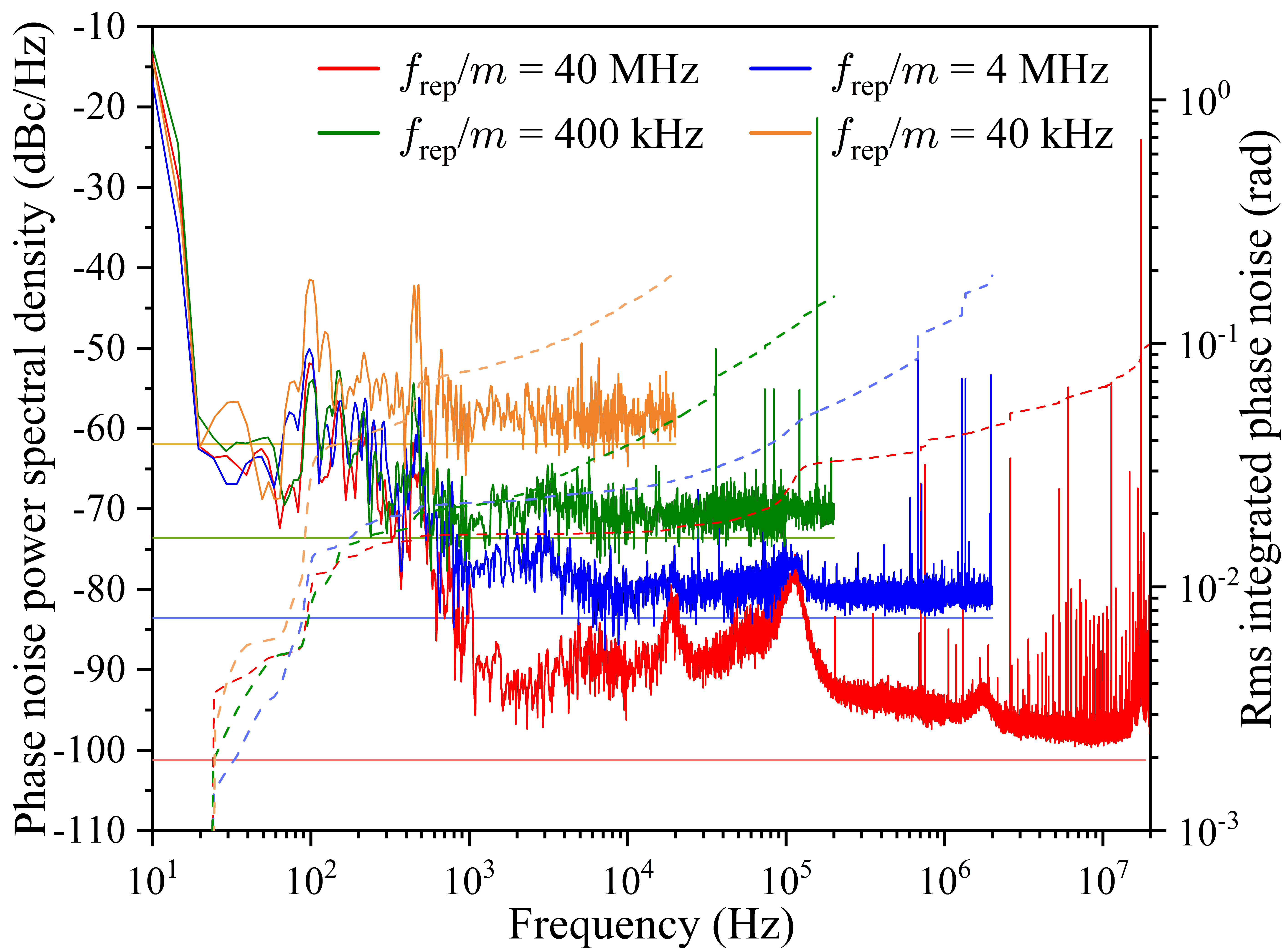}
\caption{Power spectral density (PSD) of phase noise at different repetition rates obtained by the GATOR technique (solid traces). Straight horizontal lines indicate the average of the noise floor of the measurement (only cw reference processed in the same way as the beat signal) for each PSD trace. The rms integrated phase noise is also shown (dashed lines).
The endpoint of the integration is at the Nyquist frequency, i.e. $f_{\rm rep}/2m$. Higher frequency noise contributions are aliased such that the endpoints of the dashed lines give the total phase noise. These endpoints are comparable for all the picked pulse trains. The strong peak at about 20 MHz on the 40 MHz repetition rate trace is due to a neighboring beat note.}
\label{phasenoise}
\end{figure}
%%%%%%%%%%%

The power spectral density (PSD) of the phase noise is obtained by Fourier transforming the recorded time trace, assuming that $\varphi(t)$ is small. Analyzing the heterodyne beat note allows to bring optical phase noise to the RF domain, where it can be recorded and analyzed. In a sense, the convolution of Eq.(\ref{electric_field_3}) converts the RF noise of $\tilde{\varphi}(\omega)$ into the optical domain, while heterodying brings it back into the RF domain.
The time traces contain $2.05\times10^8$ samples, and the Blackman window was applied before performing the Fourier transform. %The computed FFT is then squared and multiplied by a factor of 2 to calculate the single-sided PSD, then amplitude-normalized with respect to the value of the beat note peak analyzed by looking at its sideband spectrum.
The sideband spectrum of the Fourier transform trace around the beat frequency gives the PSD of the phase noise after normalization to the peak amplitude of the beat-note.
%\textcolor{red}{The power spectral densities obtained this way are shown in the SUPPLEMENTARY(or maybe better to insert here the ungated PSD figure? maybe without rms and floors, just pure traces).}
Since the beat signal becomes weaker when lowering the repetition rate of the comb, the phase noise PSD experiences a relative increase in the noise floor. 
To overcome this issue, the gated optical noise reduction (GATOR) technique described by Desch\^enes \textit{et al.} in Ref. \cite{Deschenes2013} was used to evaluate the data.
In the time domain, the beat note of a frequency comb with a cw laser can be understood as the pulse train sampling the cw wave, i.e. the beat note signal is available only for the duration of the pulses.
The GATOR technique strongly suppresses the noise from the cw reference laser and the detection setup by evaluating the beat signal only within narrow time windows around the comb pulses.
In the frequency domain, the GATOR method effectively averages the spectrum of the beat notes between the cw laser and several different comb modes. This increases the signal-to-noise ratio compared to beat detection with a single comb mode. We used a software implementation of GATOR by introducing temporal gate windows around each of the pulses. The width of the gate windows was set to \SI{40}{\nano\s} for all repetition rates, limited by the bandwidth of the detector.
The resulting phase noise PSDs are shown with solid lines in Fig. \ref{phasenoise}.
To evaluate the noise floor, we repeated the same measurement without comb pulses and reperformed the same analysis.
Since for traces in which the pulse separation time is lower than the gate window width the GATOR does not affect the data, we used for simplicity the ungated data to compute the \SI{40}{\MHz} of Fig. \ref{phasenoise}.
%Applying the same time gate windows on traces measured in the absence of the comb pulses, we computed the background floor for each repetition rate. To have a clearer representation of the floor level, we averaged all points of each background.
%This way, we obtained a floor level of about \SI{-103}{\dBc/\Hz}, \SI{-84}{\dBc/\Hz}, \SI{-74}{\dBc/\Hz}, and \SI{-62}{\dBc/\Hz} for \SI{40}{\MHz}, \SI{4}{\MHz}, \SI{400}{\kHz}, and \SI{40}{\kHz} phase noise PSD, respectively.

The PSD of the original frequency comb at \SI{40}{\MHz} repetition rate contains low-frequency phase noise up to about \SI{1}{\kHz} due to uncompensated environmental vibrations. Two broad noise peaks at about $\SI{20}{\kHz}$ and \SI{100}{\kHz} are servo bumps from feedback loops controlling the fast PZT actuator in the laser cavity and the AOM used for phase stabilization, respectively. The servo bump of the PZT for path length stabilization is around \SI{400}{\Hz}.
The peak at \SI{17.3}{\MHz} in the \SI{40}{\MHz} trace is due to the neighboring beat note.
Other spurious signals that we attribute to radio frequency pickups are visible at frequencies beyond \SI{100}{\kHz}.
The noise floor is determined by the amplitude noise of the cw reference laser, the noise of the photodetectors, and the electronics used in the beat detection setup. It increases for lower repetition rates due to reduced carrier power of the beat signal. The low-frequency noise below \SI{1}{\kHz} is the same for repetition rates of \SI{40}{\MHz}, \SI{4}{\MHz}, and \SI{400}{\kHz}. At a repetition rate of \SI{40}{\kHz}, some of the noise structure below 1 kHz appears 5 to 10 dB higher than other repetition frequencies, which contributes to the increase of the rms integrated phase noise of about 50~mrad.  This might be due to the different gain settings of the path length stabilization feedback. No other significant increase in phase noise was observed above the noise floor for the pulse-picked frequency combs. The FWHM linewidth of the frequency comb is limited by the resolution bandwidth of the measurement (\SI{10}{\Hz}) for all repetition rates investigated.
%Apart from the FWHM linewidth limited by the resolution bandwidth of the measurement, the power fraction-based linewidth can be defined to be a frequency range where half of the signal power is contained \cite{Schmid2019}. They are estimated to be about a few hertz for the \SI{400}{\kHz} beat signal, that is also limited by the resolution bandwidth of the measurement.

The rms phase noise integrated from \SI{20}{\Hz} to half the repetition rate $f_{\rm rep}/2m$ was calculated from the PSD spectra.
They are $\SI{100}{\milli\radian}$, $\SI{190}{\milli\radian}$, $\SI{156}{\milli\radian}$, and $\SI{195}{\milli\radian}$ for repetition rates of \SI{40}{\MHz}, \SI{4}{\MHz}, \SI{400}{\kHz}, and \SI{40}{\kHz},  respectively.
The contributions of the harmonic peaks of $f_{\rm rep}$ and the beat notes are not representing the phase noise of the comb mode and are therefore excluded from the integration. For all repetition frequencies, the measurement noise floor contributes significantly to the rms integrated phase noise.
Therefore, our measurement gives the upper limit of the phase noise of the pulse-picked comb.

The measurements presented here are based on in-loop signals. In our case, this is sufficient to determine the noise of the low repetition rate laser relative to the cw reference laser. The phase noise is not less than the measurement noise floor at any frequency. Therefore, our in-loop phase noise measurement adequately reflects the upper limit of the phase noise of our low repetition rate laser system relative to the cw reference laser. Note that the path length stabilization used here should be implemented in any application of the low repetition rate frequency comb.

%In general, it is often considered that the in-loop error signal does not reflect the noise of the signal under study because the measurement setup itself may introduce the noise, which in turn is added to the system to be stabilized. In our case, the phase noise added to the low repetition rate comb in the beat detection setup is negligible because the setup contains only a few mirrors on a stable breadboard. The feedback loop may counteract the electronic noise and bring the noise below the measurement noise floor, which in turn adds noise to the frequency comb. This is not the case in our measurement because the obtained integrated phase noise is not less than the measurement noise floor at any frequency. 

%As expected from the discussion in Sec.~\ref{theory}, the rms noise is about the same for all repetition frequencies, except \SI{40}{\MHz} where the integration range is instrumentally limited.

It is interesting to consider what limits the lowest possible repetition rate. In the laser system discussed here, the high repetition rate frequency comb is stabilized before pulse-picking using a method described in Sect. \ref{experiment}. The amplifier setup after pulse picking is prone to additional noise and more difficult to actively stabilize.
This is because the error signal for path length stabilization is taken from low-repetition-rate pulses and the available feedback bandwidth is limited to half the repetition rate.
In the time domain, the servo system needs to receive at least two pulses before it can know if the phase shift has changed.
As a result, the phase noise at Fourier frequencies significantly lower than half the repetition rate can be efficiently suppressed, while noise at higher frequencies remains unsuppressed. Note that feedback systems based on Proportional-Integral-Differential (PID) controllers require their control bandwidth to be much larger than the frequency of the noise in order to achieve a small phase delay that allows high feedback gain at lower frequencies while suppressing oscillation at higher frequencies \cite{Bechhoefer2005}. The amazing accuracy of optical frequency combs comes from their steady-state operation, i.e. a regular pulse train that finds the same environment for each pulse.
Steady-state operation is disturbed when the pulse-to-pulse interval is longer than the typical time scale of the phase noise introduced by the environment, and the disturbance cannot be actively stabilized due to limited feedback bandwidth. This problem is significant for repetition rates below 10 kHz where acoustic vibrations of many standard optical components are significant. The lowest possible repetition rate depends on the type of amplifier and the design of the setup. Our amplifiers are pumped with a cw pump source and therefore pulse-to-pulse phase variations are minimized. It also does not rely on the CPA scheme which can introduce wavelength-dependent phase fluctuations. Our demonstration shows that the compact and simple solid-state amplifiers used here introduce sufficiently low phase noise to maintain the comb structure at a repetition rate as low as 40 kHz.

To increase the feedback bandwidth significantly beyond the repetition rate after pulse picking, a probe laser with a higher repetition rate or even a continuous wave laser could be superimposed on the beamline to monitor the phase variations.
However, we expect limitations of this method because the probe beam would have a different peak intensity and spectrum. Intensity and spectrum-dependent phase shifts expected in an optical amplifier, pulse compressor, and nonlinear frequency conversion cannot be easily accounted for in this way.

%%%%%%%%%%%%%%

\section{Conclusions and future prospects}
In this paper, we have demonstrated a low repetition rate optical frequency comb based on a Yb:KYW solid-state mode-locked oscillator using an AOM pulse picker. The repetition rate is adjustable over three orders of magnitude from \SI{40}{\MHz} to \SI{40}{\kHz}.
One of the modes of the frequency comb is tightly phase-locked to a cavity-stabilized ultra-low noise cw laser by measuring the heterodyne beat note between them and providing feedback to the cavity length of the oscillator, an external AOM, and a PZT actuated mirror in the beamline.
We have characterized the phase noise of the frequency comb at repetition rates of \SI{40}{\MHz}, \SI{4}{\MHz}, \SI{400}{\kHz} and \SI{40}{\kHz} with respect to the reference cw laser.
The results confirm that a narrow linewidth comb structure is preserved even after pulse picking.
Using the power spectral density of the phase noise obtained using the GATOR technique, the integrated rms phase noise was evaluated to be \SI{195}{\milli\radian} at \SI{40}{\kHz} repetition rate.
For the first time to the best of our knowledge, we demonstrate optically-stabilized low-noise frequency combs at repetition rates as low as a few tens of kHz.

The pulse energy and average power were  \SI{200}{\nano\joule}  and \SI{8}{\milli\watt} at a repetition rate of \SI{40}{\kHz}.  A solid-state amplifier similar to the one used in this work can be added to increase the pulse energy to a >10 \SI{}{\micro\joule} level. Frequency combs with such high pulse energy and moderate average power are expected to be useful for driving high harmonic generation processes and generating XUV frequency combs \cite{Schonberg2023}. The low noise and narrow linewidth comb modes shown here indicate that low repetition rate frequency combs are a promising option for high-resolution spectroscopy at exotic wavelengths. Dual comb spectroscopy at XUV wavelengths can be an interesting application of the low repetition rate XUV frequency combs. %With the obtained integrated rms phase noise frequency conversion up to the 5th harmonic would not cause carrier collapse. It should be noted that the phase noise obtained in this work is limited by the measurement noise floor and provides an upper bound on the true phase noise. 

When direct frequency comb spectroscopy is performed using frequency combs with a low repetition rate, it may be difficult to determine the comb mode number. If the pulse picking factor is large and the uncertainty of the line center determination is small enough, the comb mode number can be unambiguously determined by repeating the measurement for different pulse picking factors.

The single-pass pulse picking scheme used in this study is inefficient for reducing the repetition rate because most of the original pulses are unused. In the future, we plan to perform pulse picking in a femtosecond buildup cavity to avoid the loss of average power \cite{Vidne2003, Jones2004}. Intracavity pulse picking will also serve as a narrow spectral filter to efficiently suppress phase and amplitude noise at frequencies greater than half the resonance width.

\section*{Funding}
This project has received funding from the European Research Council (ERC) under the European Union’s
Horizon 2020 research and innovation programme (grant agreement No. 742247). 
T.W. H\"ansch acknowledges support from the Max-Planck Foundation.

\section*{Data availability}
Data supporting the results presented in this paper are available from the authors upon reasonable request.

\section*{Disclosures}
The authors declare no conflicts of interest.

\section*{Supplemental document}
See Supplement for supporting content.

%%%%%%%%%% If using BibTeX
\bibliography{bibliography}

\newpage
\renewcommand{\theequation}{S\arabic{equation}}
\onecolumn
\section*{Supplement}
\label{Supplement1}

In Eq.(2) we assumed that the gating pulses are much longer than the laser pulses. In this supplement, we give a more rigorous derivation of the frequency comb structure after pulse picking without making this assumption. We also consider a possible timing jitter of the gating pulses. 
We show that this jitter has very little effect on the picked frequency comb, as expected from the time domain description.

We assume a rectangular shaped gating pulse
\begin{equation}
  r(t) = \left\{
         \begin{array}{ll}
           1 & -\tau/2<t<+\tau/2 \\
           0 & \, \textrm{otherwise} \\
         \end{array}
         \right.
\end{equation}
with a temporal full width $\tau$. A train of gating pulses separated in time by $mT$ is then given by
\begin{equation}
  g(t)=\sum_{q=-\infty}^{\infty} r\left(t-qmT-\tau_q\right),
\end{equation}
where we added a timing jitter of $\tau_q$ to the $q$-th gating pulse. The Fourier transform of the gating pulse train is:
\begin{equation}
  \tilde{g}(\omega)=\tau \, \mathrm{sinc}\!\left( \frac{\omega \tau}{2}\right) \, \sum_{q=-\infty}^{\infty} \, e^{-i \omega (qmT+\tau_q)}.
\end{equation}
The sinc function is defined as $\mathrm{sinc}(x)=\sin(x)/x$. The Fourier transform of the original pulse train Eq.(1) is:
\begin{equation}
  \tilde{E}(\omega) =\sqrt{\pi} \Delta t \, A \,
  \exp\!\left(-\frac{\Delta t^2}{4}\left(\omega-\omega_0\right)^2\right) \,
  \sum_{k=-\infty}^{\infty} e^{i(\omega-\omega_0)kT}.
\end{equation}
We have assumed that the carrier envelope frequency vanishes, i.e. $\omega_0kT$ is an integer multiple of $2\pi$, so the $\omega_0$ term in the sum can be dropped. The phase fluctuation $\varphi(t)$ in Eq.(1) is assumed to be zero. With the phase fluctuation $\varphi(t)$ included, the following discussion still applies to the noise less component of the spectrum $\tilde{E}_{\rm p, 0}(\omega) $.  The Fourier transform of the picked pulse train is then given by the following convolution:

\begin{eqnarray}
\tilde{E}_{\rm p}(\omega) & = & \frac{1}{2 \pi}\left(\tilde{E} \ast \tilde{g}\right)(\omega) \nonumber \\
  & \! = \! & \! \frac{\sqrt{\pi} \tau \Delta t}{T} A \sum_{n=-\infty}^{\infty}
         \exp\!\left(-\frac{\Delta t^2}{4}\left(\frac{2 \pi n}{T}-\omega_0\right)^2\right)
         \, \mathrm{sinc}\!\left(\left(\omega -\frac{2 \pi n}{T} \right)\frac{\tau}{2} \right) \nonumber \\
         & & \times \, \sum_{q=-\infty}^{\infty} e^{-i \omega qmT -i(\omega-2\pi n/T) \tau_q}.
\label{eqnnsiz}
\end{eqnarray}
Here,  the Fourier representation of the Dirac comb 
\begin{equation}
  \sum_{n=-\infty}^{\infty} \delta(x-n y) = \frac{1}{y} \sum_{p=-\infty}^{\infty} e^{i 2\pi p x/y},
\label{eq:DiracComb}
\end{equation}
was used to replace the sum over $k$ with a sum of $\delta(\omega'-2\pi n/T)$, which allows to evaluate the convolution integral. 

We will now discuss the two limiting cases that are of interest. In both cases we assume that the pulse duration is much smaller than the gating width ($\Delta t \ll \tau$). 
This is a reasonable assumption for picosecond and femtosecond pulse trains. %\footnote{For nanosecond and longer pulses, the gating width can be comparable to or even shorter than the pulse duration. The spectral bandwidth is strongly influenced by the gating width. We will not discuss this regime because such a frequency comb with a long pulse duration is not interesting for driving nonlinear processes efficiently.}

\begin{enumerate}

\item{\bf{Small timing jitter:}  $\tau_q \ll \tau$}

Using the approximation of 
\begin{eqnarray}
\mathrm{sinc}\!\left(\left(\omega -\frac{2 \pi n}{T} \right)\frac{\tau}{2} \right) e^{-i(\omega-\frac{2\pi n}{T}) \tau_q} 
&=&
\frac{i }{\left(\omega -\frac{2 \pi n}{T} \right)\tau}\left(e^{-i \left(\omega -\frac{2 \pi n}{T} \right)\left( \tau_q+\frac{\tau}{2}\right)}-e^{-i \left(\omega -\frac{2 \pi n}{T} \right)\left( \tau_q-\frac{\tau}{2}\right)}  \right) \nonumber\\
&\approx& \mathrm{sinc}\!\left(\left(\omega -\frac{2 \pi n}{T} \right)\frac{\tau}{2} \right),
\end{eqnarray}
we obtain

\begin{eqnarray}
\tilde{E}_{\rm p}(\omega) \! & \! \approx \! & \! \frac{ \sqrt{\pi} \tau \Delta t}{T} A \sum_{n=-\infty}^{\infty}
         \exp\!\left(-\frac{\Delta t^2}{4}\left(\frac{2 \pi n}{T}-\omega_0\right)^2\right)
         \, \mathrm{sinc}\!\left(\left(\omega -\frac{2 \pi n}{T} \right)\frac{\tau}{2} \right) \nonumber \\
         & & \times \, \frac{2\pi}{mT} \, \sum_{n'=-\infty}^{\infty} \delta\left(\omega-\frac{2\pi n'}{mT}\right),
         \label{eqnnonoise}
\end{eqnarray}

where we have used the Fourier representation of the Dirac comb again. The series of $\delta$-functions represents the dense frequency comb modes, and the sinc function represents the spectral intensity of the modes added by the amplitude modulation of the pulse picker. The gating pulse jitter $\tau_q$ disappears in this approximation. Exploting the finite width of the sinc-function we can approximate the argument of the Gaussian envelope as:

\begin{eqnarray}
\tilde{E}_{\rm p}(\omega) \! & \! \approx \! & \! \frac{ \sqrt{\pi} \tau \Delta t}{T} A \exp\!\left(-\frac{\Delta t^2}{4}\left(\omega-\omega_0\right)^2\right)\sum_{n=-\infty}^{\infty}
         \, \mathrm{sinc}\!\left(\left(\omega -\frac{2 \pi n}{T} \right)\frac{\tau}{2} \right) \nonumber \\
         & & \times \, \frac{2\pi}{mT} \, \sum_{n'=-\infty}^{\infty} \delta\left(\omega-\frac{2\pi n'}{mT}\right) \\
\! & \! = \! & \!  \frac{2 \pi^{3/2}\Delta t}{mT} A \exp\!\left(-\frac{\Delta t^2}{4}\left(\omega-\omega_0\right)^2\right)
  \sum_{n'=-\infty}^{\infty} \delta\left(\omega-\frac{2\pi n'}{mT}\right)  \\
&= &\mathrm{Eq.(4)}\nonumber 
\end{eqnarray}

The first line is a very good approximation as long as $\Delta t \ll \tau$, which is the regime we are interested in (optical pulse duration much shorter than the width of the gating pulses). For the sum over the sinc-function we used

\begin{equation}
 \sum_{n=-\infty}^{\infty} \mathrm{sinc}\!\left(\left(\omega -\frac{2 \pi n}{T} \right)\frac{\tau}{2} \right) =\frac{T}{\tau},
\end{equation}
which can be proven using the Poisson summation formula.
Hence the various sinc contributions to the spectral envelope add up to a constant. Remarkably, the spectral envelope of the picked frequency comb retains the Gaussian shape of the original comb. This is expected from the approximate time domain treatment in the main text. \\

Instead of taking the limit of a narrow sinc function, we can consider the opposite case of a narrow Gaussian spectral envelope. In this case, we would replace $\omega$ in the sinc function by $\omega_0$, i.e., by the central value of the (narrow) Gaussian spectral envelope, such that the gating pulses generate the spectral envelope of the comb. We mention this limit only for the sake of completeness, although it is by no means in an interesting regime for the topic of this work: The associated long pulse durations (nanoseconds) do not effectively drive nonlinear processes.

\item{\bf{Non-negligible timing jitter:}} \\

To describe the noise added by the timing jitter $\tau_q$ we consider the frequencies that contribute separately by setting
\begin{equation}
  \tau_q=\tau_0 \sin(\omega_\mathrm{noise} qmT).
\end{equation}
Using the Jacobi-Anger expansion
\begin{equation}
  e^{iz\sin(\theta)} = \sum_{k=-\infty}^{\infty} J_k(z) \, e^{ik\theta},
\end{equation}
we obtain:
\begin{eqnarray}
\tilde{E}_{\rm p}(\omega) \! & \! = \! & \! \frac{\sqrt{\pi} \tau \Delta t}{T} A \sum_{n=-\infty}^{\infty}
         \exp\!\left(-\frac{\Delta t^2}{4}\left(\frac{2 \pi n}{T}-\omega_0\right)^2\right)
         \,  \mathrm{sinc}\!\left(\left(\omega -\frac{2 \pi n}{T} \right)\frac{\tau}{2} \right) \nonumber \\
         & & \times \, \sum_{q,k=-\infty}^{\infty} J_k\left( \left(\omega-\frac{2\pi n}{T}\right)\tau_0\right)
         \,\, e^{-iq(\omega-k \omega_\mathrm{noise}) mT}.\nonumber \\
          & \! \approx \! & \! \frac{\sqrt{\pi} \tau \Delta t}{T} A 
         \exp\!\left(-\frac{\Delta t^2}{4}\left(\omega-\omega_0\right)^2\right)
         \,\nonumber \\
         & & \times \, \frac{2\pi}{mT} \sum_{n,n',k=-\infty}^{\infty} \!  \mathrm{sinc}\!\left(\left(\omega -\frac{2 \pi n}{T} \right)\frac{\tau}{2} \right)  
         J_k\left( \left(\omega-\frac{2\pi n}{T}\right)\tau_0\right)
         \,\, \delta \left(\omega-\frac{2\pi n'}{mT}-k \omega_\mathrm{noise} \right).
\end{eqnarray}

Again, we have assumed that the Gaussian spectral envelope is much broader than the sinc-function, i.e. $\Delta t \ll \tau$. In the time domain this means that the optical pulses have no tail that the gating pulses could cut away. We also used Eq.(\ref{eq:DiracComb}) again. Additional sidebands at harmonics of the noise frequency $k\omega_\mathrm{noise}$ appear in the argument of the $\delta$ function. If we consider a particular noise sideband with indices $n'$ and $k$, the remaining sum over $n$ can be evaluated using the Poisson summation formula:
\begin{equation}
\sum_{n=-\infty}^{\infty} \!  \mathrm{sinc}\!\left(\left(\omega -\frac{2 \pi n}{T} \right)\frac{\tau}{2} \right) J_k\left( \left(\omega-\frac{2\pi n}{T}\right)\tau_0\right) =0, \label{eq:BesselSinc}\\\
\end{equation}
which is valid for $k\neq 0$ when $\tau_0<\tau/2$ and $\tau_0+\tau/2<T$. These conditions mean that the edges of the gating pulses do not touch the optical pulses. The timing jitter noise is completely canceled in this case. This noise immunity is what is readily expected from the time domain picture, and it is reassuring to see it in the frequency domain as well, albeit in a much more complicated way.

\end{enumerate}

\end{document}